\documentclass[11pt,tightenlines,aps,prd,nofootinbib,superscriptaddress,eqsecnum]{revtex4}

\usepackage{amsmath,amssymb}
\usepackage[mathscr]{euscript}
\usepackage{graphicx}
\usepackage[pdftex]{color}
\usepackage[sort&compress]{natbib}
\usepackage[normalem]{ulem}  
\usepackage[colorlinks=true,linkcolor=blue,filecolor=blue,urlcolor=blue,
            citecolor=blue,pdftex=true,plainpages=false]{hyperref}

\definecolor{darkcyan}{rgb}{0, 0.5, 0.5}

\begin{document}

\title{First-order nonlinear eigenvalue problems involving functions of
a general oscillatory behavior}

\author{Javad Komijani}
\email{jkomijani@phys.ethz.ch}
\affiliation{Department of Physics, University of Tehran, Tehran 1439955961, Iran}
\affiliation{Institute for Theoretical Physics, ETH Zurich, 8093 Zurich, Switzerland}

\date{\today}

\begin{abstract}
  Eigenvalue problems arise in many areas of physics, from solving a classical
  electromagnetic problem to calculating the quantum bound states of the
  hydrogen atom. In textbooks, eigenvalue problems are defined for linear
  problems, particularly linear differential equations such as time-independent
  Schr\"odinger equations. Eigenfunctions of such problems exhibit several
  standard features independent of the form of the underlying equations.
  As discussed in Bender \emph{et al}
  [\href{http://dx.doi.org/10.1088/1751-8113/47/23/235204}{J.~Phys.~A 47,
  235204 (2014)}],
  separatrices of nonlinear differential equations share some of these features.
  In this sense, they can be considered eigenfunctions of nonlinear differential
  equations, and the quantized initial conditions that give rise to the
  separatrices can be interpreted as eigenvalues.
  We introduce a first-order nonlinear eigenvalue problem involving a general
  class of functions and obtain the large-eigenvalue limit by reducing it to a
  random walk problem on a half-line. The introduced general class of functions
  covers many special functions such as the Bessel and Airy functions, which
  are themselves solutions of second-order differential equations.
  For instance, in a special case involving the Bessel functions of the first
  kind, i.e., for $y'(x)=J_\nu(xy)$, we show that the eigenvalues asymptotically
  grow as $2^{41/42} n^{1/4}$.
  We also introduce and discuss nonlinear eigenvalue problems involving the
  reciprocal gamma and the Riemann zeta functions, which are not solutions to
  simple differential equations. With the reciprocal gamma function, i.e., for
  $y'(x)=1/\Gamma(-xy)$, we show that the $n$th eigenvalue grows factorially
  fast as $\sqrt{(1-2n)/\Gamma(r_{2n-1})}$, where $r_k$ is the $k$th root of
  the digamma function.
\end{abstract}

\maketitle

\section{Introduction}
\label{sec:intro}

In the context of stability and instability, the idea of nonlinear eigenvalue
problems was proposed first in reference \cite{Bender:2014nonlinear} for the
nonlinear first-order differential equation
\begin{equation}
  y'(x) = \cos (\pi x y), \quad y(0) = E_{n}\, ,  \label{eq:cos}
\end{equation}
where $E_n$ are the critical initial conditions that give rise to unstable
\emph{separatrix} solutions.
Putting the initial condition at the origin aside and tackling the first-order
differential equation by expanding its solutions about infinity as
$y \sim c/x + \cdots$ yields an asymptotic expansion that does not have any
arbitrary constant, while there must exist exactly one; see references
\cite[eq.~5]{Bender:2014nonlinear} and~\cite[eq.~48]{Bender:2009wx}.
It turns out the \emph{missing}, arbitrary constant lies beyond all orders.
A hyperasymptotic analysis (asymptotics beyond all orders) reveals
the structure of the expansion as well as the arbitrary
constant, and it explains the existence of separatrices~\cite{Bender:2009wx,
Bender:2014nonlinear}.
Reference~\cite{Bender:2014nonlinear} investigates the discrete spectrum of
critical initial conditions associated with the separatrices, interprets them
as the eigenvalues of the problem, and calculates the asymptotic behavior of
the eigenvalues as well as the separatrices by reducing the nonlinear problem
to a linear random-walk problem.

Various nonlinear equations appear in mathematical physics, and it would be
interesting to study them in the context of nonlinear eigenvalue problems.
Applications of this idea to the Painlev\'e equations showed that the
eigenvalues of the first, second, and fourth Painlev\'e equations
are asymptotically related to cubic, quartic, and sextic anharmonic quantum
oscillators, respectively~\cite{Bender:2015bja, Bender:2021ngq}.
Further investigations led to the introduction of a vast
class of generalized Painlev\'e equations~\cite{Bender:2019gen}.
In all cases, the nonlinear problems are reduced to linear ones at
the large-eigenvalue limit.
In this paper, we extend the program and investigate the first-order
differential equation
\begin{equation}
  y'(x) = F (x y), \quad y(0) = E , \label{eq:F}
\end{equation}
and we obtain the large order behavior of the critical initial values,
i.e., eigenvalues, for a general class of generating functions $F$ as well
as an isolated example as described below.
In the most general case, our solution involves reducing the nonlinear
problem to a random walk problem in one dimension.

Reference~\cite{Bender:2019gen} discusses the similarities between
separatrices of a nonlinear differential equation such as equation \eqref{eq:cos}
and eigenfunctions of linear time-independent Schrödinger equations,
and it clarifies the use of terminology \emph{eigenfunctions}
and \emph{eigenvalues} for nonlinear problems.
In particular, reference \cite{Bender:2019gen}
explains that eigenvalue problems are inherently unstable
because an infinitesimal change in the problem's parameters violates
the boundary conditions.
For linear problems, one can explain this instability using the Stokes
phenomenon and the Stokes multipliers.
(See reference \cite{Bender-Orszag} for a pedagogical description of the Stokes
phenomenon.)
For instance, consider the quantum harmonic oscillator
\begin{equation}
   - \psi''(x) + \frac{1}{4} x^2 \psi(x) = \left(\nu + \frac{1}{2}\right)\psi(x)
   \label{eq:parabolic:cylinder}
\end{equation}
with the boundary conditions $\psi(\pm\infty) = 0$.
This is the Weber equation, also known as the parabolic cylinder equation.
This equation has a solution in the complex plane denoted by $D_\nu(z)$
that is subdominant---vanishes exponentially fast---as $z$ tends to infinity
and $|\text{arg} z|<\pi/4$.
This special solution satisfies the vanishing boundary condition at $+\infty$,
but not necessarily the one at $-\infty$.
To find solutions that vanish at both limits, one can exploit the functional
relation
\begin{subequations}
  \begin{align}
    D_\nu(z) &= s(\nu) D_{-\nu-1}(-iz) +  e^{i\nu\pi}D_\nu(-z)\,,
    \label{eq:parabolic:D_nu} \\
    s(\nu) &= \frac{\sqrt{2\pi}}{\Gamma(-\nu)} e^{i(\nu+1)\pi/2} \,,
    \label{eq:parabolic:s}
  \end{align}
\end{subequations}
which relates subdominant solutions of the Weber equation at different regions;
see references \cite{Bender-Orszag, kawai2005algebraic} for more discussions.
The coefficient $s(\nu)$ in the above relation is called the Stokes multiplier.
Taking the boundary conditions into account,
one can argue that the eigenvalues of equation \eqref{eq:parabolic:cylinder}
are nothing but the roots of the Stokes multiplier $s(\nu)$,
which are non-negative integers, i.e., $\nu \in [0, 1, 2, \cdots]$.
For any other values, even infinitesimally different from a root, solutions of
equation \eqref{eq:parabolic:cylinder} cannot satisfy the boundary conditions.
This feature is common between the eigenvalues of linear equations
such as equation \eqref{eq:parabolic:cylinder} and the nonlinear
ones such as equation \eqref{eq:cos}.

By converting equation \eqref{eq:parabolic:cylinder} to a Riccati equation,
reference \cite{Wang:2020abc} introduces an exactly solvable nonlinear eigenvalue
problem.
That study is important because it presents a unique relationship
between a class of nonlinear eigenvalue problems and corresponding linear ones,
and it provides another justification for using
terminology \emph{eigenfunctions} and \emph{eigenvalues} for nonlinear problems.
It is noteworthy that converting a Schr\"odinger equation to a Riccati
equation lies at the heart of the WKB method.
As we discuss briefly below, in the context of the WKB method,
one can interpret a linear eigenvalue problem associated with a second-order
Schr\"odinger-type equation as a special case of a first-order nonlinear
eigenvalue problem.

It is evident from the above discussion that the eigenvalues of
equation \eqref{eq:parabolic:cylinder} grow algebraically.
This behavior is indeed another common feature between many
linear and nonlinear eigenvalue problems. For instance,
the large eigenvalues of the Schr\"odinger equation with the class of
$\mathcal{PT}$-symmetric Hamiltonians $H = \hat{p}^2 + g \hat{x}^2 (i \hat x)^{\epsilon}$
($\epsilon > 0$) grow as $n^\gamma$ with $\gamma = (2\epsilon+4)/(\epsilon+4)$,
where $\gamma$ varies between 1 and 2 depending on the value of $\epsilon$.
(This result is obtained by using the complex WKB techniques discussed in
reference \cite{Bender:1998prl}.)
Analogously, section \ref{sec:Stokes-line} presents a class of nonlinear problems
with algebraic growth of eigenvalues as $n^\gamma$, where $\gamma$ varies
between 0 and infinity.

As an example of algebraic growth of eigenvalues,
reference \cite{Bender:2014nonlinear} shows that the eigenvalues $E_n$ of
the nonlinear equation \eqref{eq:cos} grow as
\begin{equation}
   E_n \sim 2^{5/6} n^{1/2} \quad (n\to\infty). \label{eq:cos:eig:asymp}
\end{equation}
An alternative proof of this asymptotic behavior is given in
reference \cite{Kerr:2014alternative}, and an attempt toward exact WKB analysis of
the problem is presented in reference \cite{Shigaki:2019abc}.
In a similar study, reference \cite{Bender:2019gen} investigates a special case of
equation \eqref{eq:F} with the generating function $F$ set to the Bessel function of
the first kind and order 0 and finds numerically that
\begin{equation}
   E_n \sim A n^{1/4} \quad (n\to\infty), \label{eq:J_nu:eig:asymptotic}
\end{equation}
with $A \approx \frac{35}{18}$.%
\footnote{The numerical analysis of reference \cite{Bender:2019gen}
    yielded a value for the constant $A$ with ambiguity in its third digit,
    which agrees with $\frac{35}{18} \approx 1.94444$
    as well as $2^{41/42} \approx 1.96726$ within the uncertainties.
    This is in contrast to the numerical precision in 
    reference \cite{Bender:2014nonlinear} that achieved an accuracy of one part in
    $10^{10}$ and led to a reliable conjecture that the overall coefficient in
    equation \eqref{eq:cos:eig:asymp} is indeed $2^{5/6}$,
    which was confirmed analytically.
    The reason for such a difference in accuracy (with double-precision
    arithmetic) is discussed in section \ref{sec:Stokes-line}.
}
In this paper, we derive this relation analytically and obtain $A=2^{41/42}$.
Moreover, we show this asymptotic behavior is valid for all Bessel functions of
the first kind and order $\nu\ge0$.
The proof that we provide here is a generalization of the method developed
in reference \cite{Bender:2014nonlinear} to tackle equation \eqref{eq:cos},
and it is applicable for a general class of functions $F$ that asymptotically
oscillate as
\begin{equation}
  F(x) \sim a x^\alpha \cos\left(b x^\beta + \varphi\right)
  \label{eq:intro:F}
\end{equation}
as the argument of the function approaches infinity.
This is indeed the asymptotic behavior of solutions of ordinary differential
equations such as the Bessel and Airy functions 
on their Stokes lines.%
\footnote{Here we use the convention of reference \cite{Bender-Orszag} to define
    Stokes lines.
}
We also extend the study to a couple of functions that are not solutions to
ordinary differential equations, such as the \emph{reciprocal} gamma function,
which is proportional to the Stokes multiplier of the parabolic cylinder
equation, and the Riemann zeta function.

The rest of the paper is organized as follows.
In the next section, we discuss nonlinear eigenvalue
problems with a class of generating functions $F$ with
asymptotic behavior specified in equation \eqref{eq:intro:F},
and we calculate the large-eigenvalue limit analytically.
Numerical solutions of special cases of $F$, namely the Bessel and Airy
functions, are also presented in the next section.
In section \ref{sec:reciprocal-gamma}, we solve a similar problem 
involving the reciprocal gamma function.
Concluding remarks, including a discussion on the zeta function as a generating
function and relation between nonlinear and linear eigenvalue problems
in the context of the WKB method, are presented in section \ref{sec:conclusion}.

\section{Models with asymptotically oscillatory functions}
\label{sec:Stokes-line}

\subsection{Problem definition}

In this section, we take into account a general class of generating
functions $F$ that satisfy the asymptotic relation
\begin{equation}
  F(x) \sim a x^\alpha \cos\left(b x^\beta + \varphi\right)
  \label{eq:def:F:asymp}
\end{equation}
as $x\to\infty$. Many functions, including the Bessel and Airy functions,
satisfy this asymptotic form on their Stokes lines:
\begin{equation}
  J_\nu(x) \sim \sqrt{\frac{2}{\pi x}} \cos\left(x - \frac{2\nu+1}{4}\pi \right)
  \label{eq:Bessel:asymp}
\end{equation}
as $x\to\infty$ and
\begin{equation}
  \text{Ai}(x) \sim \frac{1}{\sqrt{\pi} (-x)^{1/4}} 
  \cos\left(\frac{2}{3}(-x)^{3/2} - \frac{\pi}{4}\right)
  \label{eq:Airy:asymp}
\end{equation}
as $x\to-\infty$.
With $F$ from such a general class of functions,
we define the nonlinear eigenvalue problem
\begin{equation}
  y'(x) = F \left(x y\right), \quad y(0) = E , \label{eq:F:repeat}
\end{equation}
and determine the initial conditions that give rise to separatrix solutions
as $x\to\infty$.

\begin{figure}
  \includegraphics[width=0.49\textwidth]{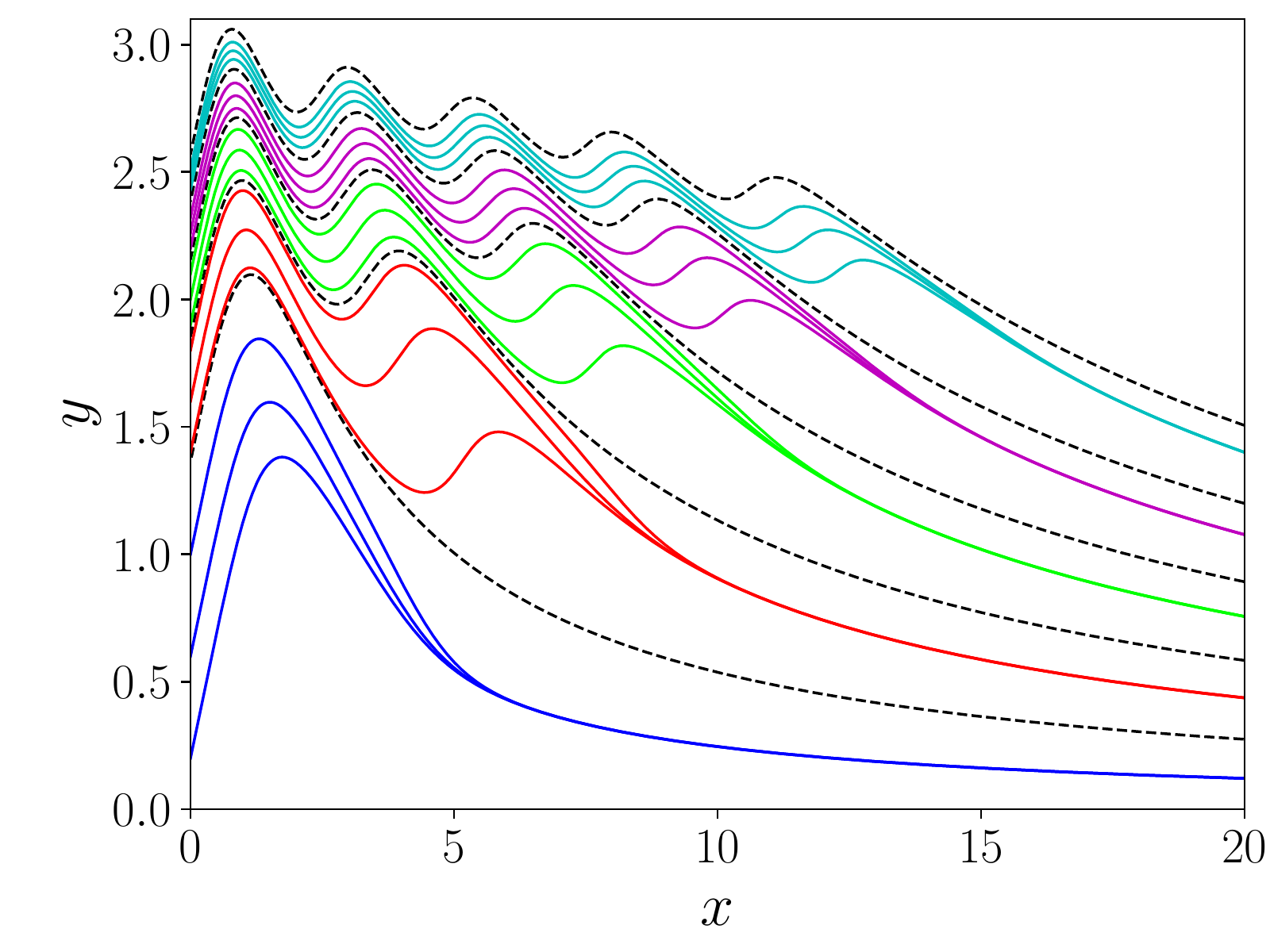}
  \includegraphics[width=0.49\textwidth]{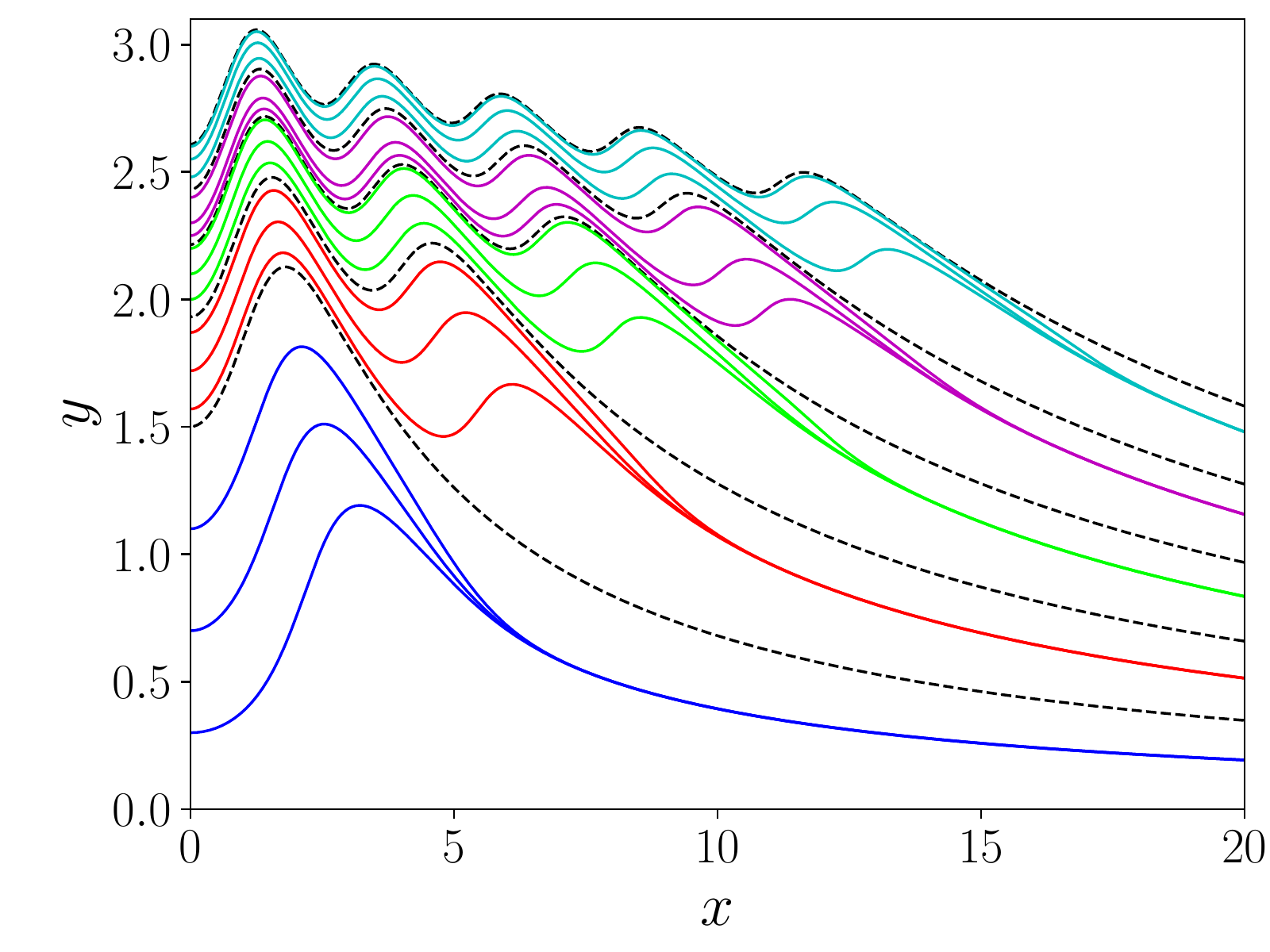}
  \caption{Solutions of equation \eqref{eq:J_nu} with $\nu=0$ (left) and $\nu=1$
  (right) for twenty initial values. Five of the initial values are tuned
  to the critical values corresponding to the separatrix solutions
  shown by dashed curves.
  }
  \label{fig:Bessel:few_eigen_unscaled}
\end{figure}

Before tackling the problem in its general form,
we briefly explore a special case of the problem with the Bessel functions
from a numerical point of view. 
Figure~\ref{fig:Bessel:few_eigen_unscaled} illustrates solutions of
\begin{equation}
   y'(x) = J_\nu (x y), \quad x\ge 0, \label{eq:J_nu}
\end{equation}
with $\nu=0$ (left) and $\nu=1$ (right) for twenty initial values $y(0)$.
Among the initial values of each panel, five of them are tuned to critical
values (eigenvalues) corresponding to the separatrix solutions shown by
dashed curves.
One can observe that when the initial condition is slightly different from an
eigenvalue, the solution veers away from the corresponding separatrix and gets
attracted to a nearby stable asymptotic solution.
A hyperasymptotic analysis, similar to the one presented for
equation \eqref{eq:cos} in reference \cite{Bender:2014nonlinear}, is required to
understand this phenomenon.
Because the solutions are qualitatively very similar to the solutions of
equation \eqref{eq:cos}, we refer the reader to reference \cite{Bender:2014nonlinear}
for a detailed explanation of this phenomenon.

Let us briefly review the properties of the $n$th separatrix
in the left (right) panel of figure \ref{fig:Bessel:few_eigen_unscaled}.
As $x$ increases from 0, $y(x)$ oscillates with exactly $n$ maxima
and then decays to 0 monotonically as $x\to\infty$.
This behavior resembles a quantum wave function that oscillates in the so-called
classically \emph{allowed} region and decays in the classically \emph{forbidden}
region.
Inspired by high-energy semiclassical calculation of eigenfunctions and
eigenvalues in quantum mechanics using the WKB method,
we introduce a method to study the asymptotic behavior of the separatrices
shown in figure \ref{fig:Bessel:few_eigen_unscaled} and corresponding eigenvalues.
The method that we present here is a generalization of the one developed in
reference \cite{Bender:2014nonlinear} to tackle equation \eqref{eq:cos}
and is applicable not only for problems involving the Bessel functions
but also for the general case defined in equation \eqref{eq:F:repeat}.

\subsection{Asymptotic behavior}

The WKB method is a powerful tool to calculate the asymptotic behavior of
eigenvalues in quantum mechanics.
To this end, the WKB method provides different asymptotic expansions
for the wave function, which hold in their respective regions of validity,
and then joins them together to obtain a global solution by matching
the solutions in neighboring regions.
The matching is done in the so-called \emph{turning-point} regions
and puts constraints on possible solutions.
In this part, we calculate the asymptotic behavior of
eigenvalues in equation \eqref{eq:F:repeat} following the same strategy of
the WKB method. We also employ the quantum-mechanics terminology of classically
\emph{allowed} and \emph{forbidden} regions as well as \emph{turning points}.

We assume $a$ and $b$ in equation \eqref{eq:def:F:asymp} are real, positive
numbers, and we restrict the domain and range of the solutions to $x\ge0$ and
$y\ge0$, respectively. Unless otherwise stated, we assume $\beta>0$.
Under some general conditions on $F$ at the vicinity of origin, the structure
of separatrices is then similar to what we observed in figure
\ref{fig:Bessel:few_eigen_unscaled}.

To tackle this problem, we use the change of variables
\begin{subequations}
  \label{eq:general:xy2tz}
  \begin{align}
     y &= \sqrt{a} \left(\frac{\lambda}{b}\right)^\gamma z \,,\\
     x &= \frac{1}{\sqrt{a}} 
          \left(\frac{\lambda}{b}\right)^{\frac{1}{\beta}-\gamma} t \,,
  \end{align}
\end{subequations}
with $\gamma = (1+\alpha)/(2\beta)$
so that equation \eqref{eq:F:repeat} asymptotically reads
\begin{equation}
  \frac{dz}{dt} \sim (t z)^\alpha \cos \left(\lambda (t z)^\beta + \varphi \right),
  \label{eq:F:asymp}
\end{equation}
as $\lambda\to\infty$ for a non-vanishing $t z(t)$.
We also use the parametrization $\lambda = ((2n-\frac{1}{2})\pi-\varphi)$
such that an integer value $n$ corresponds to the ($2n$)th zero of the cosine
function and the $n$th eigenvalue of the problem.

The asymptotic solution is simple in the forbidden region $t > 1$:
\begin{equation}
  z(t) = \frac{1}{t}
         \left[1 - \frac{1}{\beta\lambda} \arcsin\left(\frac{1}{t^2}\right)
               + \text{O}\left(\frac{1}{\lambda^2}\right)
         \right]
  \label{eq:sol:t_g_1}
\end{equation}
as $\lambda\to\infty$.
Note that the change of variables that we used puts the turning point of
the problem at $t=1$ and results in a solution that approaches unity as
$\lambda$ approaches infinity;
i.e., $z(1) = 1$ at the infinite-$\lambda$ limit.
We use this result as the boundary (matching) condition of the solution
at $t < 1$.

To obtain the solution in the allowed region $t<1$, we multiply
the differential equation~\eqref{eq:F:asymp} by $(z+tz')z^{-2\alpha}$,
and we write it as
\begin{align}
  &\frac{1}{2-2\alpha} \frac{d}{dt} z^{2-2\alpha}(t)
    + \frac{1}{2} t^{1+2\alpha} 
    \left(1+\cos\left( 2 \lambda (t z)^\beta + 2\varphi\right)\right)\nonumber \\
  & \sim \frac{1}{\lambda} \left(\frac{t}{z}\right)^{\alpha}
    \frac{(t z)^{1-\beta}}{\beta}
    \frac{d}{dt} \sin \left(\lambda (t z)^\beta + \varphi\right)
    \quad (\lambda\to\infty)\,.
    \label{eq:F:mult}
\end{align}
To obtain this relation, we replaced $[z'(t)]^2$ by
equation \eqref{eq:F:asymp} 
and used the double-angle formula for the cosine function.
Integrating equation \eqref{eq:F:mult} from $t_0$ to $t$, we obtain:
\begin{align}
  \label{eq:NEP:F:solution}
  \frac{z^{2-2\alpha}(t) - z^{2-2\alpha}(t_0)}{1-\alpha}
  + \frac{t^{2+2\alpha} - t_0^{2+2\alpha}}{2+2\alpha}
  + \eta(t;t_0)
  &= \text{O}(1/\lambda)\quad (\lambda\to\infty)\,,
\end{align}
where
\begin{equation}
  \eta(t;t_0) = \int_{t_0}^t ds s^{1+2\alpha}
  \cos\left(2 \lambda (s z(s))^\beta + 2\varphi\right)
  \label{eq:F:def:eta} .
\end{equation}
Note that to obtain the right-hand side of equation \eqref{eq:NEP:F:solution},
one can use integration by parts and show that
\begin{align}
  & \int_{t_0}^t ds \frac{1}{\lambda} \left(\frac{s}{z(s)}\right)^{\alpha}
    \frac{(s z(s))^{1-\beta}}{\beta}
    \frac{d}{ds} \sin \left(\lambda (s z(s))^\beta + \varphi\right) \nonumber \\
  & = \frac{1}{\beta \lambda} \times
    \Biggl\{ \left. \frac{s^{\alpha+1-\beta}}{(z(s))^{\alpha-1+\beta}}
    \sin \left(\lambda (s z(s))^\beta + \varphi\right)\right|_{t_0}^t \nonumber \\
  & \quad\qquad - \int_{t_0}^t ds
    \frac{d}{ds} \left( \frac{s^{\alpha+1-\beta}}{(z(s))^{\alpha-1+\beta}}\right)
    \sin \left(\lambda (s z(s))^\beta + \varphi\right)
   \Biggr\} ,
   \label{eq:F:NLO:0}
\end{align}
which remains of order $1/\lambda$ as $\lambda\to\infty$.

Solving equation \eqref{eq:NEP:F:solution} is not trivial, even in the leading
order. The right-hand side of equation \eqref{eq:NEP:F:solution} vanishes at
the infinite limit of $\lambda$.
The left-hand side, on the contrary, is not easy because it involves
$\eta(t;t_0)$, which is an integral of a complicated, rapidly varying function.
We are facing a multiple-scale problem, and because of its nonlinear nature,
we cannot exploit well-known methods like the WKB method to solve the problem.
To calculate $\eta(t;t_0)$, we use a method initially developed in reference
\cite{Bender:2014nonlinear} and generalize it to fit the current problem.
The starting point is to define an infinite set of moments as
\begin{equation}
  A_{n,k}(t;t_0) \equiv \int_{t_0}^t ds\, s^{1+2\alpha} 
  \cos\left( n \lambda (s z(s))^\beta + n\varphi\right)
  \left(\frac{s^{1+\alpha}} {z^{1-\alpha}(s)}\right)^{k},
  \label{eq:def:Ank}
\end{equation}
and note that $\eta(t;t_0) = A_{2,0}(t;t_0)$.
These moments are overwhelmingly complicated,
but they satisfy a simple, linear difference equation for large $\lambda$:
\begin{equation} \label{eq:Ank:difference}
   A_{n,k}(t;t_0)
   = -\frac{1}{2} A_{n-1,k+1} (t; t_0) - \frac{1}{2} A_{n+1,k+1}(t; t_0)\, .
\end{equation}
To obtain this equation, we multiply the integrand of the integral in
\eqref{eq:def:Ank} by
\begin{equation}
  \frac{z(s) + s z'(s)}{z(s)} - \frac{s z'(s)}{z(s)}
\end{equation}
and then evaluate the first part of the resulting integral by parts
and show that it is negligible as $\lambda\to\infty$ if $t_0$ and $t$ are
not greater than unity.
In the second part of the integral, we replace $z'(t)$ by
equation \eqref{eq:F:asymp} and use the trigonometric identity
\begin{equation}
  \cos(na)\cos(a) = \frac{1}{2}\cos((n+1)a) + \frac{1}{2}\cos((n-1)a)\, .
\end{equation}

Let us use $\eta_\infty(t; t_0)$ to denote the infinite-$\lambda$
limit of $\eta(t; t_0)$.
We now exploit the linear difference equation~\eqref{eq:Ank:difference}
to calculate $\eta_\infty(t; t_0)$.
By repeated use of the difference equation,
one can expand $\eta_\infty(t; t_0)$ as the series
\begin{equation}
  \eta_\infty(t;t_0) = \sum_{p=0}^\infty \alpha_{1,2p+1} A_{1,2p+1}(t),
  \label{eq:eta_infty:expanded}
\end{equation}
where the coefficients $\alpha_{n, k}$ are determined by a one-dimensional
random-walk process in which random walkers move left or right with equal
probability but become static when they reach $n=1$.
The coefficients can be found in exact form.
We refer the reader to reference \cite{Bender:2014nonlinear}
for details, and we reproduce the result here:
\begin{equation}
  \alpha_{1,2p+1} = \frac{\Gamma(p+1/2)}{\Gamma(-1/2)(p+1)!}\, .
\end{equation}

Plugging the coefficients in equation \eqref{eq:eta_infty:expanded},
we obtain a series that remarkably can be summed in closed form: 
\begin{align}
  \eta_\infty(t;t_0) &= \lim_{\lambda\to\infty}
  \sum_{p=0}^\infty \frac{\Gamma(p+1/2)}{\Gamma(-1/2)(p+1)!}
  \int_{t_0}^t ds\, s^{1+2\alpha}  z'(s) \left(sz(s)\right)^{-\alpha}
  \left(\frac{s^{2+2\alpha}} {z^{2-2\alpha}(s)}\right)^{p+1/2}
  \nonumber \\
  &= \lim_{\lambda\to\infty} \int_{t_0}^t ds z'(s) z^{1-2\alpha}(s)
  \left(\sqrt{1-\frac{s^{2+2\alpha}} {z^{2-2\alpha}(s)}} - 1\right).
\end{align}
The final result is valid for $t_0$ and $t$ not larger than unity.
Interestingly, there is no explicit reference to $\lambda$ in this expression,
and we can safely pass to the limit as $\lambda\to\infty$.
In this limit, the function $z(t)$, which is rapidly oscillatory,
approaches the function $z_\infty(t)$, which is smooth and not oscillatory.
The function $z_\infty(t)$ obeys
\begin{align}
  \label{eq:NEP:J0:z_infty}
  \frac{z_\infty^{2-2\alpha}(t) - z_\infty^{2-2\alpha}(t_0)}{1-\alpha}
  + \frac{t^{2+2\alpha} - t_0^{2+2\alpha}}{2+2\alpha}
  + \eta_\infty(t;t_0) = 0\,.
\end{align}
We differentiate the above integral equation with respect to $t$ to obtain
an elementary differential equation:
\begin{align}
  & z_\infty'(t) z_\infty^{1-2\alpha}(t)
  \left(\sqrt{1-\frac{t^{2+2\alpha}} {z_\infty^{2-2\alpha}(t)}} + 1\right)
  + t^{1+2\alpha} = 0 .
\end{align}
A change of variables
as $z_\infty^{1-\alpha}(t) = t^{1+\alpha} u(t)$
can easily solve this problem.
The result reads
\begin{align}
  \left(z_\infty^{1-\alpha}(t)
      + \frac{\alpha-1}{2} \sqrt{z_\infty^{2-2\alpha}(t) - t^{2+2\alpha}}\right)^2
  \left(z_\infty^{1-\alpha}(t)
      + \sqrt{z_\infty^{2-2\alpha}(t) - t^{2+2\alpha}}\right)^{1-\alpha}
  = 1,
  \label{eq:F:z_infty:solution}
\end{align}
where the constant on the right-hand side is obtained by matching
the solution at the turning point with equation \eqref{eq:sol:t_g_1},
i.e., by imposing the condition $z_\infty(1) = 1$.
This concludes our derivation of $z_\infty(t)$.

We now discuss the behavior of $z_\infty(t)$ in the vicinity of the origin.
Note that our result for $z_\infty(t)$ depends only on $\alpha$,
and there are three cases depending on the value of $\alpha$:
\begin{itemize}
  \item If $\alpha>-1$, the $t^{2+2\alpha}$ terms vanish at $t=0$.
    Consequently, $z_\infty(0)$ remains finite and reads
   \begin{equation}
     z_\infty(0) = \left(\frac{2^{1+\alpha}}{(1+\alpha)^2}\right)^{\frac{1}{(1-\alpha)(3-\alpha)}} .
     \label{eq:F:z_infty:0}
   \end{equation}
 \item If $\alpha<-1$, the $t^{2+2\alpha}$ terms diverges at $t=0$.
   Therefore, as $t\to0$, we have
   \begin{equation}
     z_\infty(t) \sim
     \left(\frac{(1-\alpha) t^{1+\alpha}}{\sqrt{(1-\alpha)^2 -4}} \right)^{\frac{1}{1-\alpha}} .
     \label{eq:F:z_infty:t-to-0}
   \end{equation}
 \item If $\alpha=-1$, as $t\to0$, we obtain
   \begin{equation}
     z_\infty(t) \sim \sqrt[4]{-2 \ln t} \, .
     \label{eq:F:z_infty:t-to-0:alpha=-1}
   \end{equation}
\end{itemize}
We conclude that it is only for $\alpha>-1$ that one can define eigenvalues at
$t=0$.%
\footnote{For $\alpha\le-1$ one can define eigenvalues at $t=\tau>0$.
For instance, when $\alpha=-1$,
this leads to eigenvalues that grow like $\sqrt[4]{\ln n}$ as $n\to\infty$.}
Using equation \eqref{eq:general:xy2tz}, we find the asymptotic behavior
of the eigenvalues (for $\alpha>-1$):
\begin{equation}
  E_n \sim A n^\gamma\quad (n\to\infty),
  \label{eq:F:eig:asymp}
\end{equation}
where $\gamma = (1+\alpha)/(2\beta)$ and
\begin{align}
 A = \sqrt{a} \left(\frac{2\pi}{b}\right)^\gamma
 \left(\frac{2^{1+\alpha}}{(1+\alpha)^2}\right)^{\frac{1}{(1-\alpha)(3-\alpha)}} .
\label{eq:F:eig:asymp:A}
\end{align}
This concludes the principal asymptotic analysis of the
eigenvalues of equation \eqref{eq:F:repeat}.

\subsection{Special cases: Bessel and Airy functions}

In the previous part, we calculated the asymptotic behavior of
the eigenvalues and eigenfunctions for problems involving functions
of general oscillatory behavior.
The derived results, namely equations \eqref{eq:F:z_infty:solution},
\eqref{eq:F:eig:asymp}, and \eqref{eq:F:eig:asymp:A},
are valid for all $\beta>0$ and $1+\alpha>0$.
The problems specified in equations \eqref{eq:cos} and~\eqref{eq:J_nu},
with the cosine and Bessel functions, respectively,
are special cases of the problem we solved.

Let us now explore equation \eqref{eq:J_nu} in the light of the derived results.
Performing the change of variables \eqref{eq:general:xy2tz},
equation \eqref{eq:J_nu} reads
\begin{equation}
  \frac{dz}{dt} = \sqrt{\frac{\pi\lambda}{2}} J_\nu \left(\lambda t z(t)\right).
  \label{eq:J_nu:lambda}
\end{equation}
Then, in the limit of large eigenvalues, $z(t)$ approaches $z_\infty(t)$,
which is $1/t$ for $t \ge 1$ and
\begin{align}
  \left(4 \sqrt{z_\infty^3(t)} - 3 \sqrt{z_\infty^3(t) - t}\right)^4
  \left(\sqrt{z_\infty^3(t)} + \sqrt{z_\infty^3(t) - t}\right)^3
  &= 2^8  \label{eq:z_infty:solution}
\end{align}
for $t<1$, and the eigenvalues grow as
\begin{align}
   E_n & \sim 2^{41/42} n^{1/4}\quad (n\to\infty)\, .
\end{align}
This asymptotic relations yields the overall constant in
equation \eqref{eq:J_nu:eig:asymptotic}: $A=2^{41/42}$.

\begin{figure}
  \includegraphics[width=0.49\textwidth]{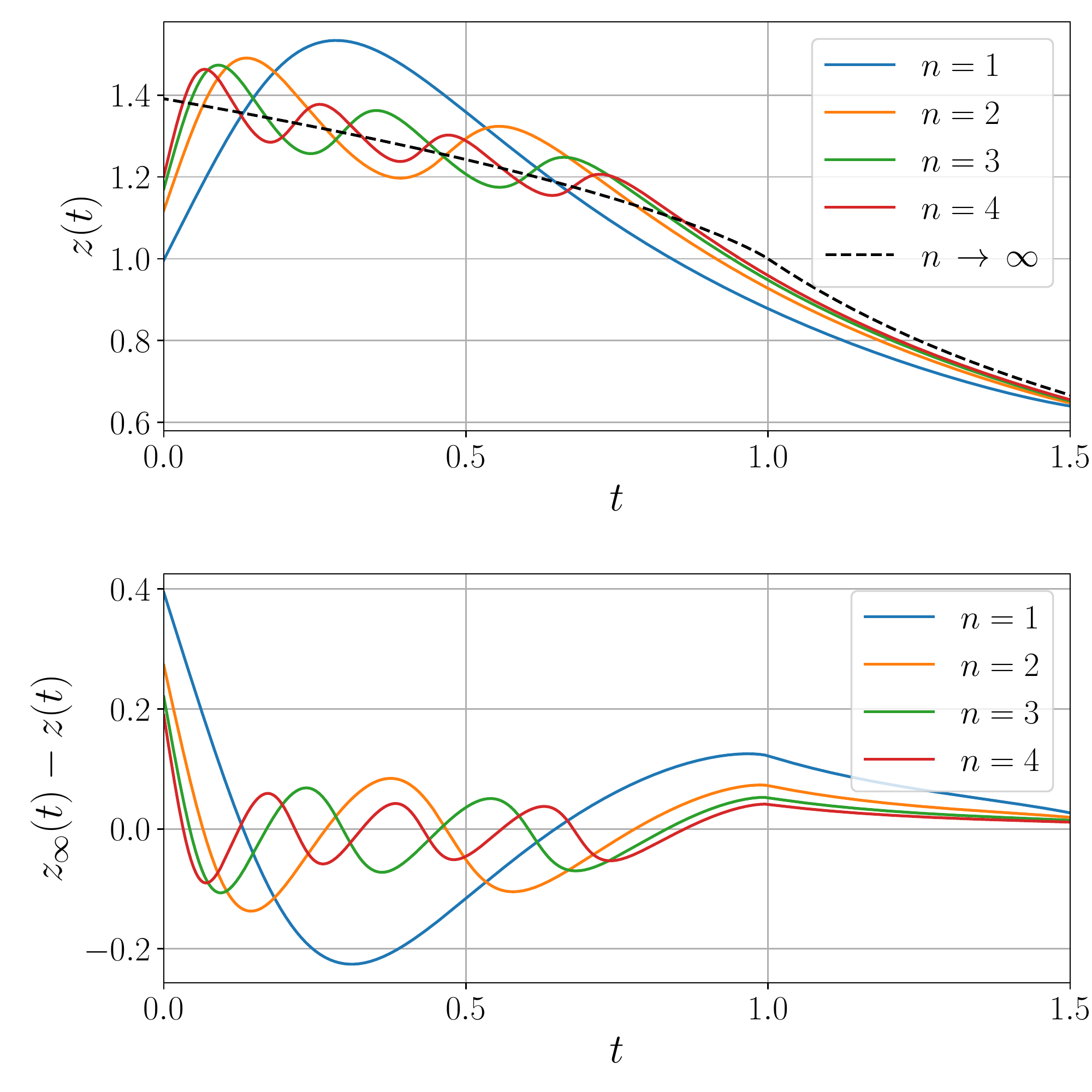}
  \includegraphics[width=0.49\textwidth]{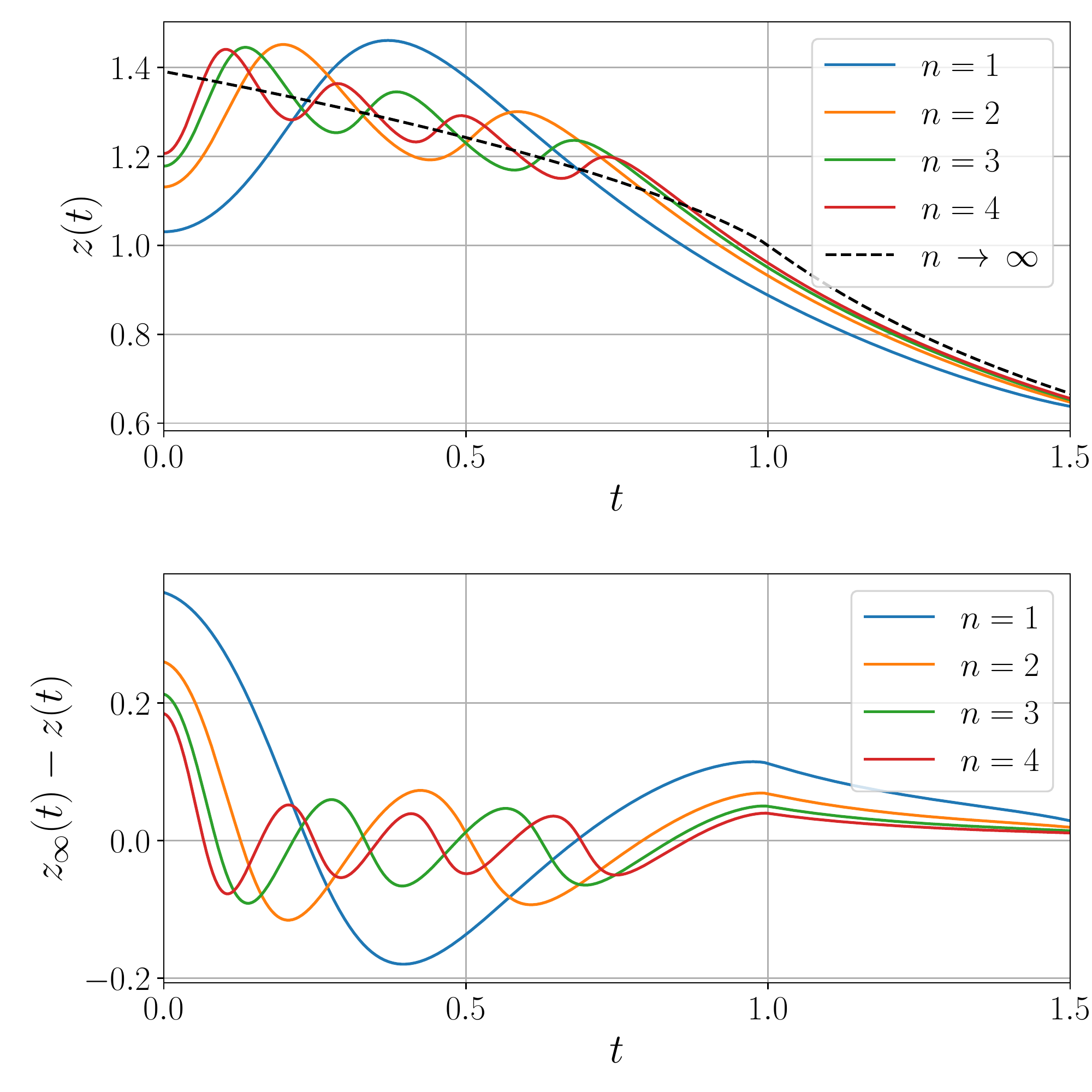}
  \caption{Upper left: the solid curves are the first four eigensolution
   $z(t)$ of equation \eqref{eq:J_nu:lambda} with $\nu=0$ corresponding to the
   Bessel function of order 0. The dashed curve is the large-$n$ limit curve
   $z_\infty(t)$ given in equation \eqref{eq:z_infty:solution}.
   Lower left: differences between the solid curves and the dashed curve.
   Upper and lower right: similar to the upper and lower left panels,
   respectively, but for the Bessel function of order 1.
  }
  \label{fig:few_eigen}
\end{figure}

\begin{figure}
  \includegraphics[width=0.49\textwidth]{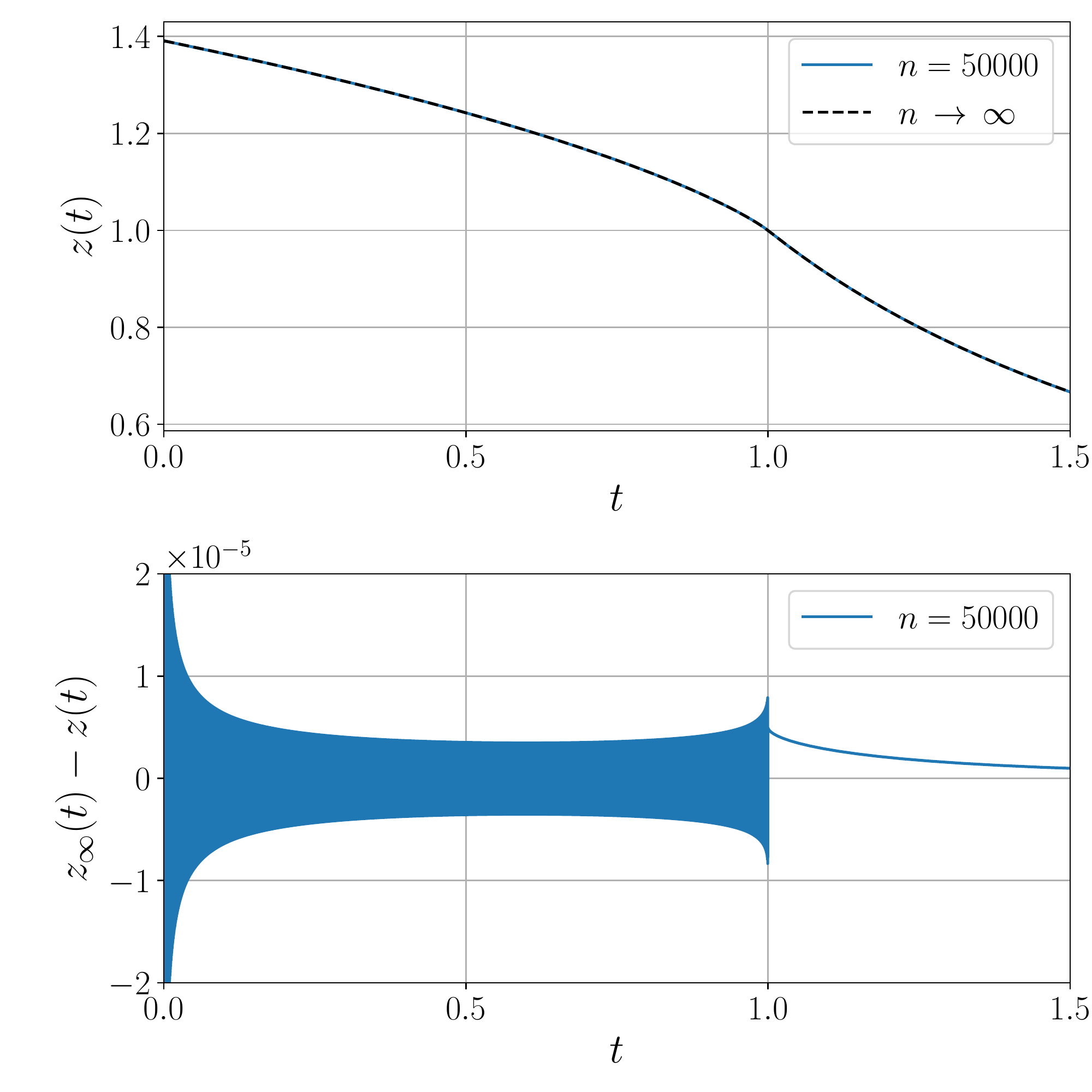}
  \includegraphics[width=0.49\textwidth]{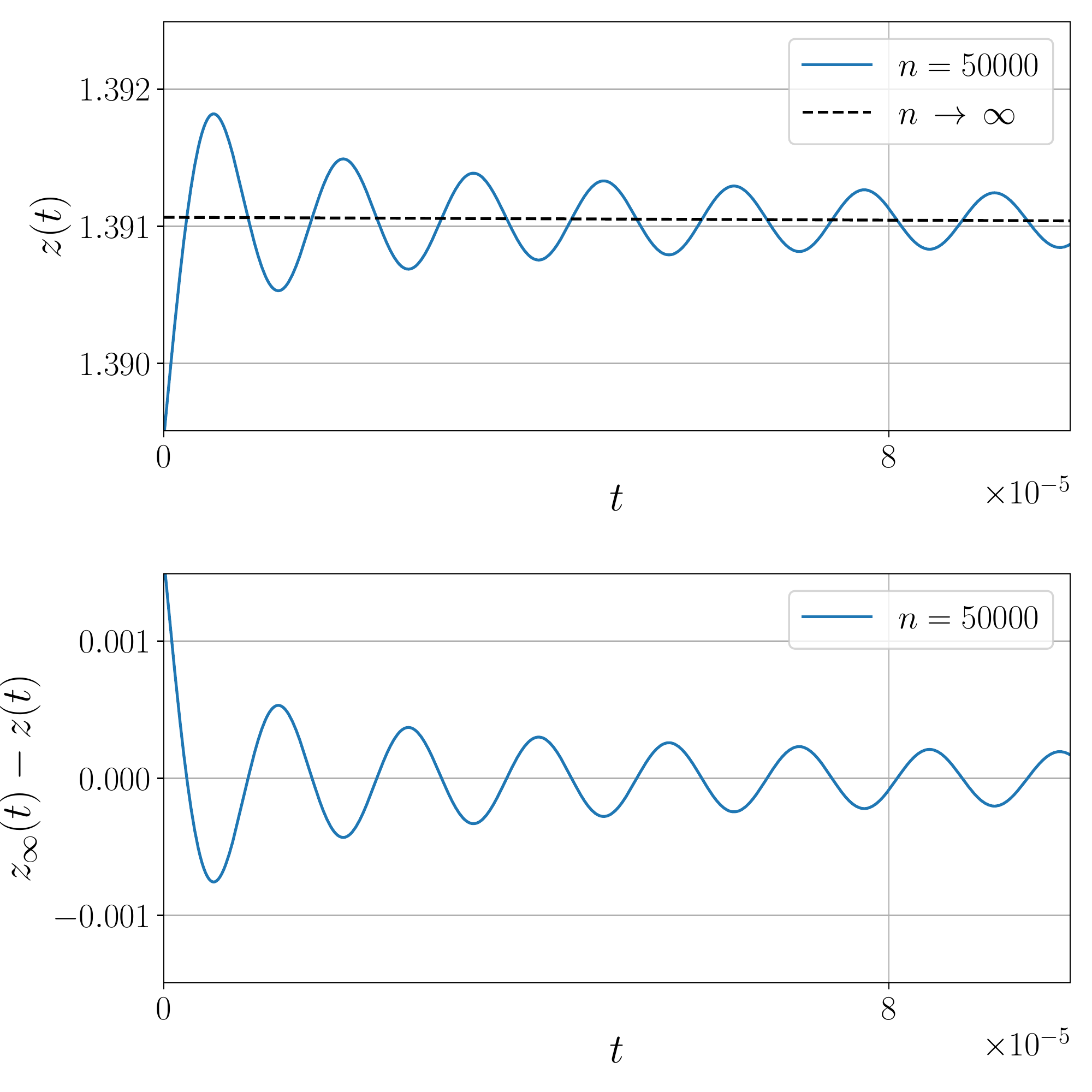}
  \caption{Upper left: the solid line is the $n = 50000$ eigensolution to
    equation \eqref{eq:J_nu:lambda} with $\nu=0$,
    and the dashed curve is the large-$n$ limit curve $z_\infty(t)$ given in
    equation \eqref{eq:z_infty:solution}.
    Lower left: difference between the solid and dashed curves of the upper left
    panel. The difference is highly oscillatory and the bulk of the oscillations
    is of the order $10^{-5}$.
    Upper and lower right: similar to the upper and lower left panels,
    respectively, but zoomed in to a region with $t$ of size $1/n$ or smaller.
  }
  \label{fig:eig50000}
\end{figure}

We now numerically compare the eigensolutions of equation \eqref{eq:J_nu:lambda}
and the large-$\lambda$ limit function $z_\infty(t)$.
Figure~\ref{fig:few_eigen} illustrates the first four eigensolutions of $z(t)$
to equation \eqref{eq:J_nu:lambda} with $\nu=0$ (upper left) and $\nu=1$ (upper
right). These eigensolutions have one, two, three, and four maxima, respectively.
They oscillate about the large-$\lambda$ limit curve $z_\infty(t)$ shown by a
dashed curve, and as $n$ increases, the amplitude of oscillations decreases.
The lower panels in figure \ref{fig:few_eigen} show the difference between
the large-$\lambda$ limit curve and the eigensolutions plotted on the upper panels.

Figure \ref{fig:eig50000} shows the $n=50000$ eigensolution to
equation \eqref{eq:J_nu:lambda} with $\nu=0$.
The difference between this eigensolution and the large-$\lambda$ limit curve
$z_\infty(t)$ is not visible in the upper left panel because the amplitude of
oscillations is tiny.
The lower-left panel shows that the \emph{envelope} modulating the rapidly
oscillating part is of order $1/n$, namely $10^{-5}$.
The envelope slowly increases to order $1/\sqrt{n}$, namely $10^{-3}$,
as $t$ approaches zero; zoomed in the right panels of figure \ref{fig:eig50000}.
(We discuss below the size of the envelope analytically.)
The $1/\sqrt{n}$ scaling of the envelope at the vicinity of the origin
indicates that the next-to-leading order corrections to
the eigenvalues $E_n$ are of size $1/\sqrt{n}$.
Therefore, one needs to go to very high values of $n$ to extract the overall
coefficient of the asymptotic behavior of eigenvalues, i.e., to obtain $A$
in equation \eqref{eq:J_nu:eig:asymptotic}.
Moreover, the Richardson extrapolation cannot work well to study the eigenvalues
because the central assumption in the Richardson extrapolation is that
the corrections to the leading term are of order $1/n$.

\begin{figure}
  \includegraphics[width=0.49\textwidth]{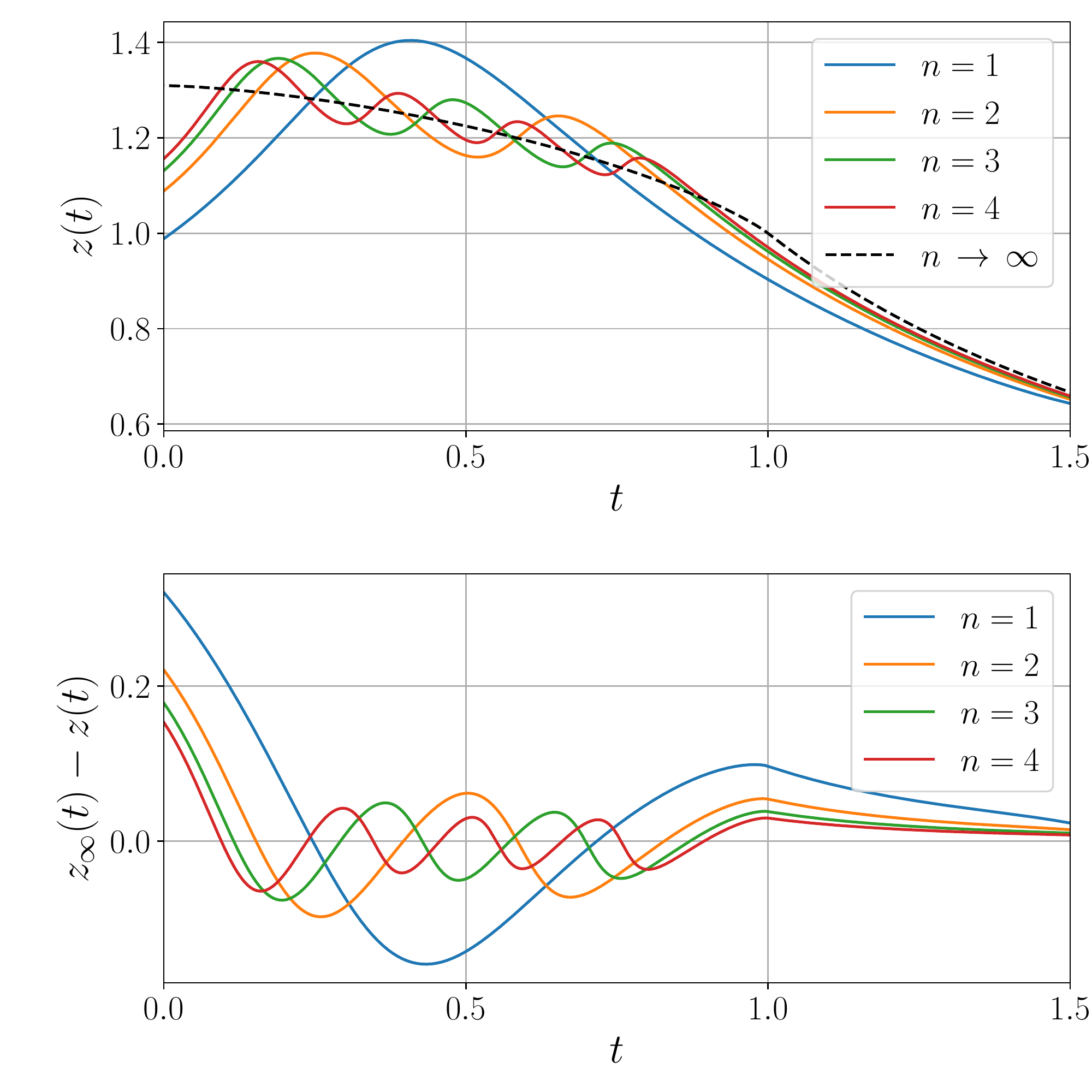}
  \includegraphics[width=0.49\textwidth]{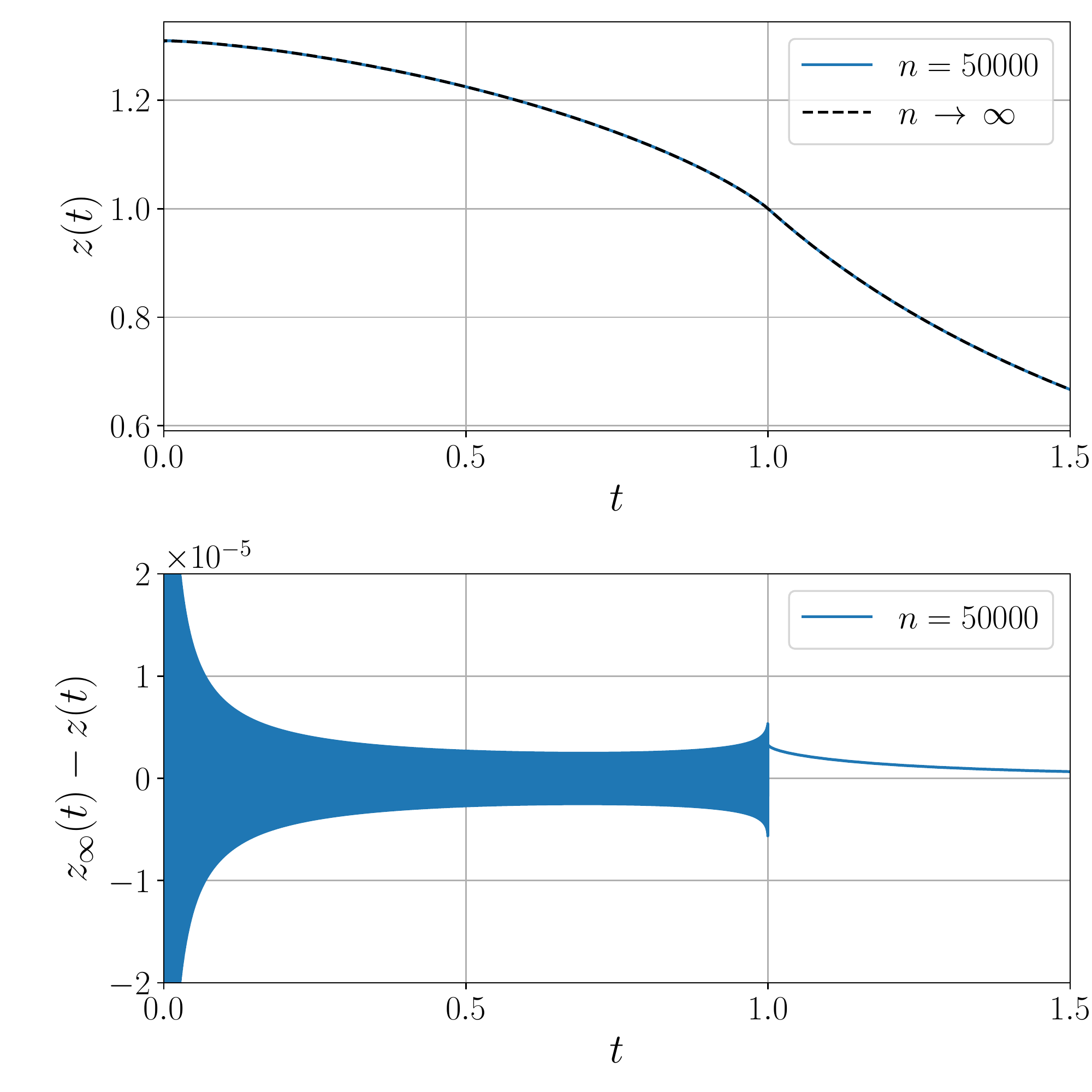}
  \caption{Left and right panels: similar to the left panels of
    figures \ref{fig:few_eigen}
    and~\ref{fig:eig50000}, respectively, but for the Airy function.
  }
  \label{fig:Airy}
\end{figure}

Another interesting case involves the Airy function on its Stokes line:
\begin{equation}
  y'(x) = \text{Ai}(-xy), \quad x \ge 0, \label{eq:Airy}
\end{equation}
with initial condition $y(0) = E$.
Note that the Airy function can be written in terms of the modified Bessel function
\begin{equation}
  \text{Ai}(x) = \frac{1}{\pi} \sqrt{\frac{x}{3}} K_{1/3}\left(\frac{2}{3}x^{3/2}\right) ,
\end{equation}
and it obeys the asymptotic relations in equation \eqref{eq:Airy:asymp}
as $x\to -\infty$.
From equation \eqref{eq:F:eig:asymp}, it is evident that the eigenvalues
of this problem behave asymptotically as $B n^{1/4}$,
similar to the Bessel function case but with a different multiplicative constant.
Figure~\ref{fig:Airy} illustrates the first four and the $n=50000$ (scaled)
eigensolutions for the Airy function.

\subsection{Further remarks}

We end the discussion of this section with a few remarks.

The numerical solutions illustrated in figures
\ref{fig:Bessel:few_eigen_unscaled}, \ref{fig:few_eigen},
and the left panels of \ref{fig:Airy} are calculated using the \emph{odeint}
function from the \emph{integrate} package in \emph{scipy},
and the ones in figure \ref{fig:eig50000} and the right panels of figure
\ref{fig:Airy} are calculated using an adaptive RK4 method.
For precise determination of the separatrix curves, which are unstable and
sensitive to numerical round-off errors for increasing $t$,
we calculate them backward from large values of $t$ down to the origin.
(Note that instability depends on the direction of integration.)

The asymptotic results presented in this section are obtained in the
large-eigenvalue limit of the problem, ignoring all terms that vanish in this
limit.
To calculate the envelope modulating the rapidly oscillating part in the lower
panels of figures \ref{fig:eig50000} and~\ref{fig:Airy}, one needs to include
next-to-leading order terms. Without discussing it in detail, we point out that
the envelope can be derived from equation \eqref{eq:F:NLO:0}:
\begin{equation}
  z_\text{env}(t) \sim \frac{1}{\beta \lambda}  t^{1+\alpha-\beta} z_\infty^{\alpha-\beta}(t) 
  \label{eq:envelope}
\end{equation}
as $t\to0$. This relation indicates that the difference $z(0) - z_\infty(0)$
vanishes like $\lambda^{-2\gamma}$ when $t$ is of order $\lambda^{-1/\beta}$.
One then concludes that, for both the Bessel and Airy functions,
the envelopes grow like $1/\sqrt{n}$ when $t$ is of order $1/\lambda$.
This conclusion agrees with the numerical solutions shown in
figures \ref{fig:eig50000} and~\ref{fig:Airy}.

So far, we assumed $\beta>0$, but the general results given in
equations \eqref{eq:F:eig:asymp} and \eqref{eq:F:eig:asymp:A}
are valid for $\beta<0$ too.
However, note that the structure of eigensolutions for $\beta<0$ are different
from those of the $\beta>0$ cases.
For instance, the number of maxima of the eigensolutions is not finite
when $\beta<0$ because equation \eqref{eq:def:F:asymp} highly oscillates as 
$x$ approaches zero.

Finally, we point out that $z_\infty(t)$ in equation \eqref{eq:F:z_infty:solution}
approaches unity as $\alpha$ approaches infinity.
As discussed in the next section, this limit is identical to the asymptotic
limit of a nonlinear eigenvalue problem involving the reciprocal gamma function.

\section{A Model with the reciprocal gamma function}
\label{sec:reciprocal-gamma}

In this section, we employ the reciprocal gamma function to define a nonlinear
eigenvalue problem:
\begin{equation}
  y'(x) = \frac{1}{\Gamma(-x y)}, \quad x\ge 0 , \label{eq:rgamma}
\end{equation}
with initial condition $y(0) = E$.
We show that the eigenvalues of this problem behave as
\begin{equation}
  E_n \sim \sqrt{-\frac{2n-1}{\Gamma(r_{2n-1})}} \label{eq:rgamma:eig:asymptotic}
  \quad(n\to\infty),
\end{equation}
where $r_{2n-1}$ is the $(2n-1)$th root of the digamma function,
\begin{equation}
  r_\lambda \approx -\lambda 
       + \frac{1}{\pi} \arctan\left(\frac{\pi}{\log(\lambda+1/8)}\right) \, .
  \label{eq:digamma:root}
\end{equation}

\begin{figure}
  \includegraphics[width=0.5\textwidth]{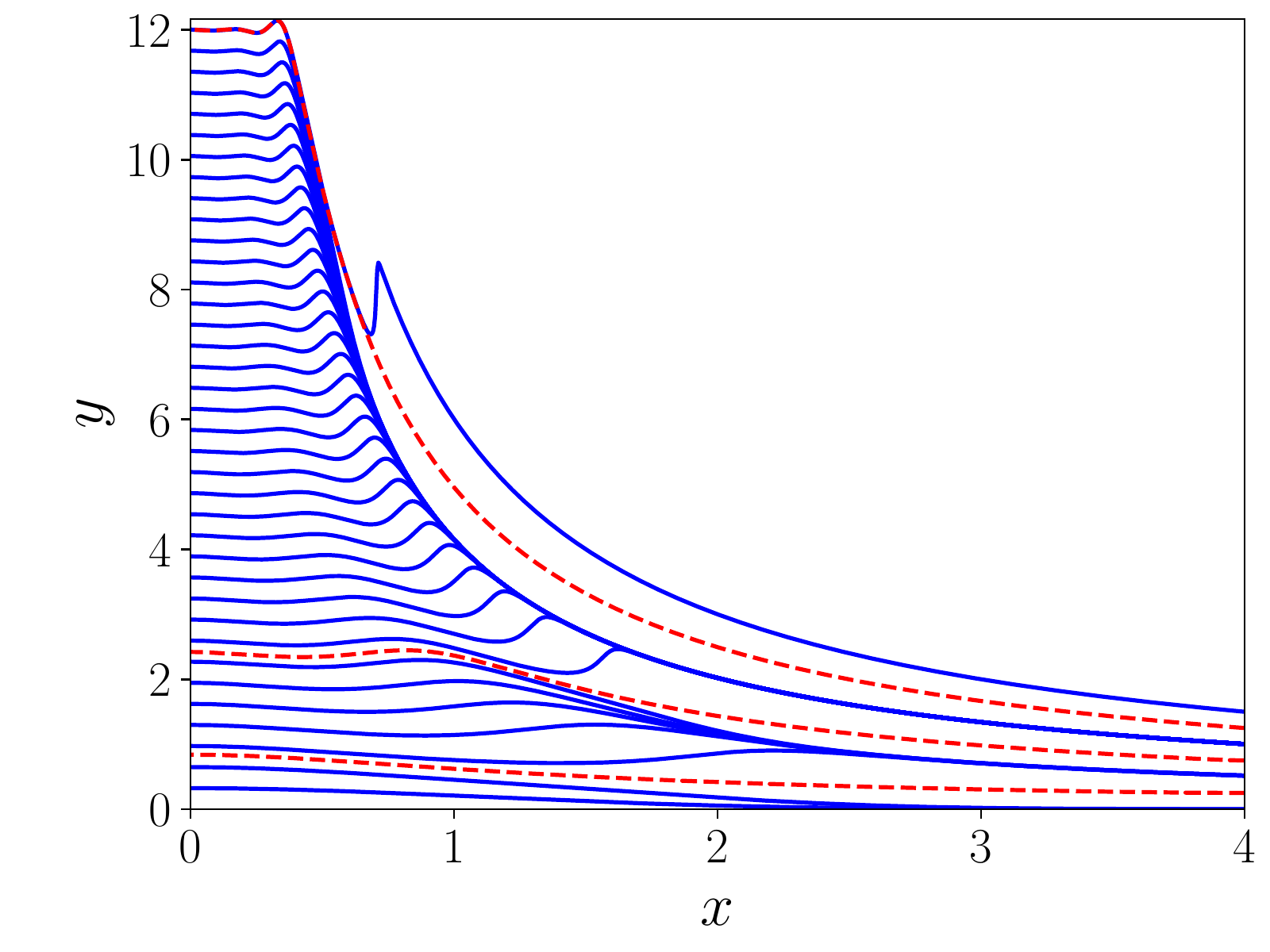}
  \caption{Numerical solutions of equation \eqref{eq:rgamma} for several initial
    values $y(0)$. The red dashed curves are the first three eigensolutions.
  }
  \label{fig:rgamma:raw}
\end{figure}

Figure \ref{fig:rgamma:raw} illustrates solutions of equation \eqref{eq:rgamma}
for several initial values $y(0)$, including the first three eigenvalues.
Like the previous examples, $y(x)$ oscillates in an allowed region
as $x$ increases from 0 and smoothly decreases in a forbidden region.
At large $x$, $y(x)$ asymptotically behaves as $c/x$, where $c = 2n-1$ for
the $n$th separatrix solution.
This behavior can be verified using the identity
\begin{equation}
  \frac{1}{\Gamma(-x y)} = -\frac{1}{\pi} \sin(\pi x y) \Gamma(1+x y)\, .
\end{equation}
The eigenvalues corresponding to the separatrices grow factorially,
as indicated by figure \ref{fig:rgamma:raw}.
For instance, the tenth and twentieth eigenvalues are 
$5.50\times10^8$ and $2.86\times 10^{23}$, respectively.
They can be compared to
$4.98\times10^8$ and $2.68\times 10^{23}$ obtained
from equation \eqref{eq:rgamma:eig:asymptotic}.
Because of the factorial growth,
double-precision arithmetic cannot handle large eigenvalues.

To obtain the large-eigenvalue limit, we employ a change of variables as
\begin{subequations}
  \label{eq:rgamma:xy2tz}
  \begin{align}
    x &= \sqrt{\lambda/\xi(\lambda)}\, t \,,\\
    y &= \sqrt{\lambda\, \xi(\lambda)}\,  z \,,
  \end{align}
\end{subequations}
where $\lambda = 2n - 1$ and $\xi(\lambda)$ is a function of $\lambda$ that
will be fixed shortly.
With this change of variables, equation \eqref{eq:rgamma} reads
\begin{equation}
  \xi(\lambda)\,\frac{dz}{dt} 
  = \frac{1}{\Gamma\left(- \lambda  t z\right)}.
  \label{eq:rgamma:change}
\end{equation}
To have a well-defined limit as $\lambda\to\infty$,
one can argue that $\xi(\lambda)$ should be
\begin{equation}
  \xi(\lambda) = \frac{-1}{{\Gamma(r_\lambda)}} \xi_0 
  \label{eq:rgamma:xi},
\end{equation}
where $r_\lambda$ is the $(2n-1)$th root of the digamma function
and $\xi_0$ is a constant or any function that approaches a constant
at the large-$\lambda$ limit.
We set $\xi_0 = 1$ and argue below that this choice corresponds to setting
the turning point of the problem to $t=1$.

To obtain the asymptotic solution of equation \eqref{eq:rgamma:xi}
in the forbidden region $t>1$, we start from the following parametrization
\begin{equation}
  t z(t) = 1 - \frac{\epsilon(t)}{\lambda \log(\lambda)}\, .
\end{equation}
We then show that as $\lambda\to\infty$, $\epsilon(t)$ satisfies
\begin{equation}
  \epsilon(t) e^{-\epsilon(t)} = \frac{\xi_0}{t^2} e^{-1} \,;
\end{equation}
$\epsilon(t) = -W_0(-\xi_0/(e t^2))$, where $W_0$ is the Lambert $W$ function
on its principal branch.%
\footnote{See reference \cite{Corless:1996lambertw} for the definition and
    properties of the Lambert $W$ function.
    In particular, note that the Lambert $W$ function has two real branches:
    $W_0(x)$ denotes the branch satisfying $-1 < W(x)$, which is called
    the \emph{principal} branch, and $W_{-1}(x)$ denotes the branch satisfying
    $W(x)\le -1$. It is noteworthy that the Lambert $W$ function appears in many
    problems in physics. Here are some examples:
    in the double-well Dirac delta potential~\cite{Corless:1996lambertw},
    in the study of the renormalon divergence in the pole mass of a quark
    \cite[eq.~3.15]{Komijani:2017vep},
    and in the QCD running coupling $\alpha_\text{g}(\mu)$ in the
    \emph{geometric} scheme.
    For the latter, see equation (2.20) in reference \cite{Brambilla:2017mrs},
    which (after correcting for typos and using $\beta_0$ and $\beta_1$ to
    denote the first two coefficients of the beta function) can be written as
    \begin{equation}
        \frac{-\beta_0}{\beta_1 \alpha_\text{g}(\mu)} =
        W_{-1}\left(-e^{-1}
        \left(\frac{\Lambda_\text{g}}{\mu}\right)^{2\beta_0^2/\beta_1}\right)
        \nonumber
    \end{equation}
    in a setting with asymptotic freedom
    ($\alpha_g(\mu)\to 0$ as $\mu \to \infty$)
    and positive $\beta_1$. Here, $\Lambda_\text{g}$ is the critical scale of
    the running coupling corresponding to the branch point of $W_{-1}(x)$ at
    $x = -e^{-1}$.
    See reference \cite[eq.~7]{Wu:2018cmb} for a similar scheme.
    The Lambert $W$ function also appears in the study of the nontrivial zeros
    of the zeta function.
}  
The critical point of $W_0$ determines the turning point of the problem:
$t_\text{turning} = \sqrt{\xi_0}$.
As we wish to put the turning point at unity, we set $\xi_0=1$.
This choice indicates $z(t)$ approaches unity as $\lambda$ approaches infinity;
i.e., $z(1) = 1$ at the infinite-$\lambda$ limit.
We exploit this result as the boundary (matching) condition for the solution
in the allowed region $t < 1$, which can be derived easily because one can argue
that $\frac{dz}{dt}$ vanished for $t<1$ at the infinite-$\lambda$ limit.
Taking the boundary condition at the turning point into account, we conclude
that $z(t)$ approaches to
\begin{align}
 \label{eq:rgamma:z_infty:solution}
 z_\infty(t) = \begin{cases}
                1\qquad   & t \le 1 \\
                1/t\qquad & t  > 1
               \end{cases}
\end{align}
as $\lambda\to\infty$.
This result, in turn, yields the asymptotic behavior that we already announced
in equation \eqref{eq:rgamma:eig:asymptotic}.

\begin{figure}
  \includegraphics[width=0.49\textwidth]{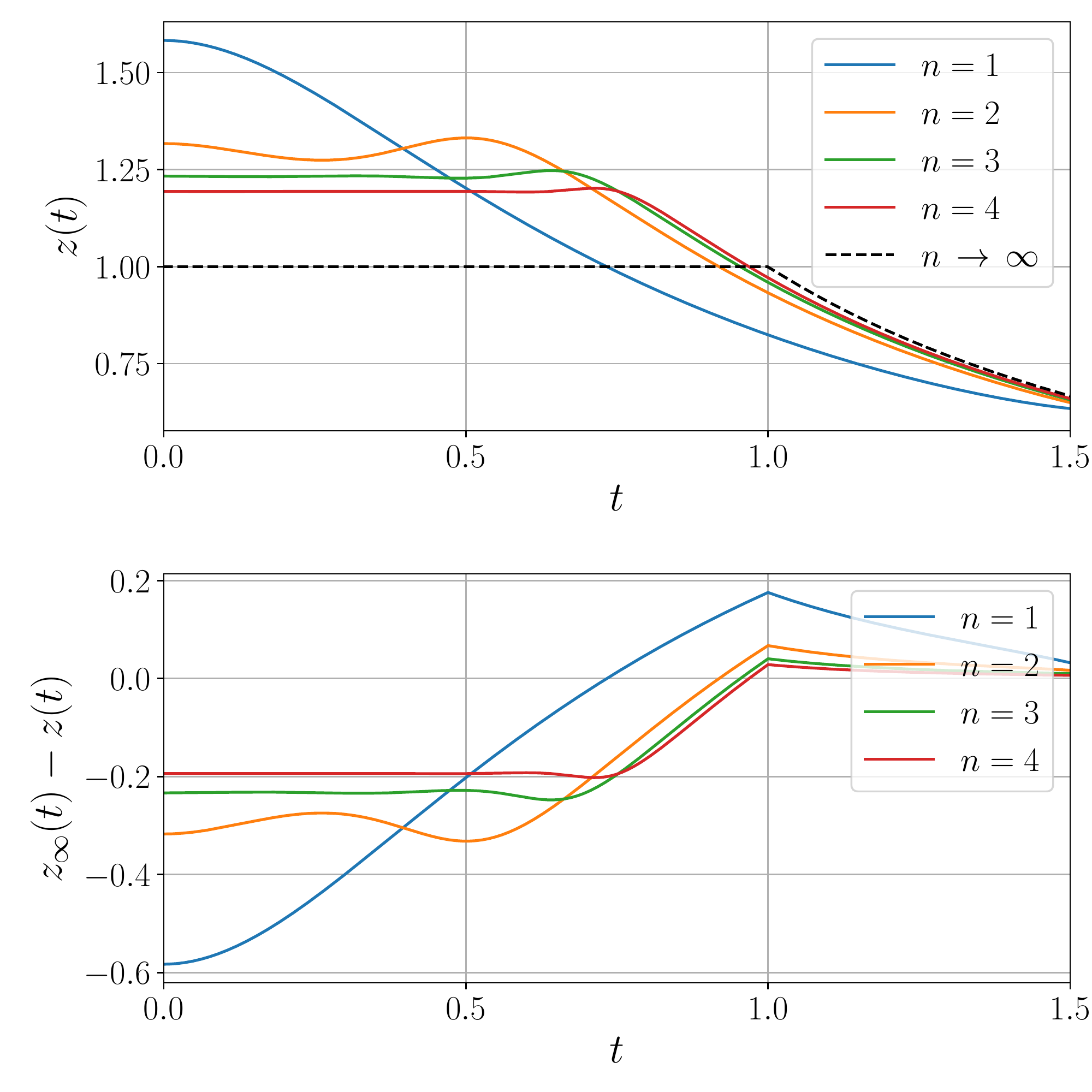}
  \includegraphics[width=0.49\textwidth]{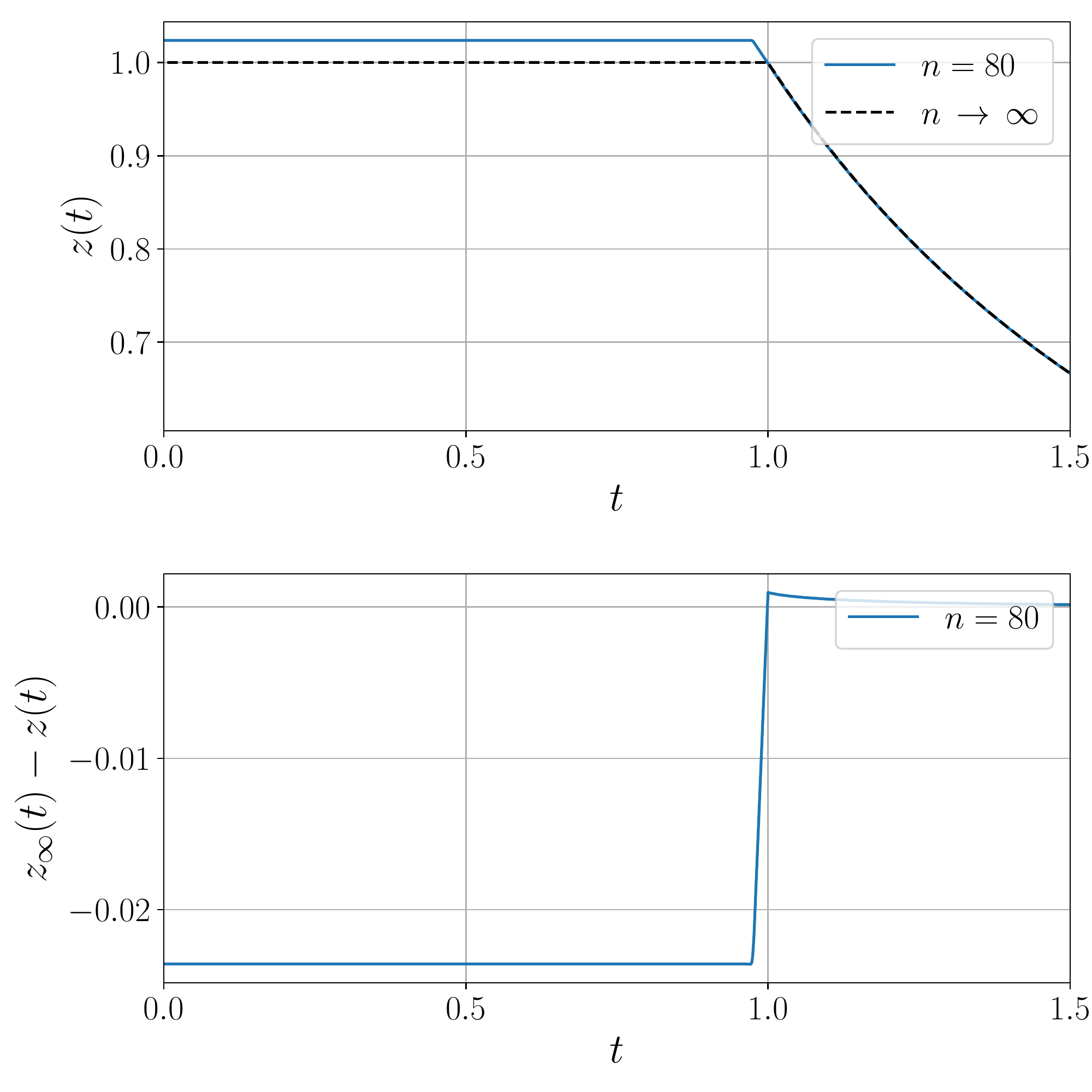}
  \caption{Similar to figure \ref{fig:Airy} but for the reciprocal gamma function.
  The right panel shows the scaled eigensolution corresponding to the 80th eigenvalue.
  }
  \label{fig:rgamma}
\end{figure}

Figure~\ref{fig:rgamma} illustrates the first four eigensolutions of $z(t)$ to
equation \eqref{eq:rgamma:change} (left panel) as well as the 80th eigensolution
(right panel). The solutions oscillate when $t<1$, but there is a bias compared
to the limit curve $z_\infty(t)$, unlike other examples discussed above.

\section{Summary and concluding remarks}
\label{sec:conclusion}

In this paper, we studied a class of first-order nonlinear eigenvalue problems
with generating functions that asymptotically behave as
$a x^\alpha \cos\left(b x^\beta + \varphi\right)$ as $x\to\infty$.
This asymptotic behavior is standard among special functions that are
solutions of ordinary differential equations on their Stokes lines.
Extending the technique developed in reference \cite{Bender:2014nonlinear},
we introduced a method to study the asymptotic behavior of large
eigenvalues of this nonlinear problem.
We can compare our method with the WKB method, which provides a way to calculate
this limit for linear eigenvalue problems of Schr\"odinger-type equations.
Consider the linear time-independent Schrödinger equation on the infinite domain
$ -\infty < x < \infty$
\begin{equation} \label{eq:Schrodinger}
  - \psi''(x) + \eta^2 \left(V(x) - E\right) \psi(x) = 0,
  \quad \psi(\pm \infty) = 0 ,
\end{equation}
where $\eta = 1/\hbar$ and $V(x)$ rises at $\pm\infty$.
The WKB method constructs solutions of the form
\begin{equation}
  \psi(x) = \exp\left(\eta \int^x S(t; \eta) dt\right) ,
\end{equation}
where $S(x; \eta)$ satisfies the Riccati equation
\begin{equation}
  \eta^{-1} S'(x; \eta) = V(x) - E - S^2(x;\eta).
  \label{eq:Schrodinger2Riccati}
\end{equation}
One can study this Riccati equation 
in the context of nonlinear eigenvalue problems.
Solving this equation makes it clear that the eigenfunctions of
the Schr\"odinger equation are closely related to the eigenfunctions of
the Riccati equation.
It is straightforward%
\footnote{Expanding $S(x; \eta)$ in inverse powers of $\eta$,
    one can show that the odd terms can be written in terms of the even terms as
    \begin{equation*}
        S_\text{odd} = -\frac{1}{2\eta}\frac{d}{dx}\log S_\text{even}
    \end{equation*}
    and the even terms satisfy
    \begin{equation*}
        \sqrt{S_\text{even}} \frac{d^2}{dx^2}\bigl(\frac{1}{\sqrt{S_\text{even}}}\bigr)
        = \eta^2 (V(x) - E - S_\text{even}^2) \, .
    \end{equation*}
    Interesting, $-S_\text{even}$ is also a solution of the above equation.
    The rest of the calculation is straightforward:
    starting from $S_\text{even} = \sqrt{V(x) - E} + \text{O}(\eta^{-2})$
    one can obtain all higher-order terms recursively.
    Therefore, the two independent solutions of the Schr\"odinger equation
    \eqref{eq:Schrodinger} read
    \begin{equation}\label{WKB_large-eta}
        \psi_{\pm}(x) = \frac{1}{\sqrt{S_\text{even}}}
            e^{\pm \eta \int^x S_\text{even} dx'}\;. 
    \end{equation}
    The only difficulty is that the resulting expansion is asymptotic
    with a vanishing radius of convergence. That is why the traditional WKB
    method is useful only at high energies.
    The \emph{exact} WKB~\cite{Voros:1983abc, Silverstone:1985wkb} method
    circumvents this issue by exploiting the Borel sum to tame the divergence.
    See reference \cite{kawai2005algebraic} for a brief review of the subject.
}  
to obtain the solution of the Riccati equation as an expansion in inverse powers
of $\eta$. On the contrary, for the nonlinear eigenvalue problem studied here,
it is not easy to accomplish such a mission even in the leading order.
Here we could obtain the leading term by reducing the nonlinear problem
to a linear random walk problem that can be solved exactly.
From a different point of view, the method we developed here can be considered
an extension of the WKB method tailored for our nonlinear problem.
We believe the exact WKB analysis of this problem can open a new area of
research. Reference~\cite{Shigaki:2019abc} presents such an attempt toward
exact WKB analysis of the nonlinear eigenvalue problem studied in reference
\cite{Bender:2014nonlinear},
which is only a special case of the problem studied here.
On the other hand, the method we developed here might be helpful in extending
the WKB method to nonlinear problems.

The Stokes multipliers of linear differential equations provide another class
of generating functions for first-order nonlinear eigenvalue problems.
In this paper, we investigated the reciprocal gamma function
and worked out its large-eigenvalue limit.
Another interesting example is the Riemann zeta function $\zeta(z)$.
According to the Riemann hypothesis, the nontrivial zeros of $\zeta(z)$
lie on its critical line, and there is a conjecture that the nontrivial zeros
are related to eigenvalues of a specific Hamiltonian;
see reference \cite{Bender:2016wob} and references there.
Instead of the Riemann zeta function itself, it is easier to use the Riemann
xi-function $\xi(z)$ to define a nonlinear eigenvalue problem
because it is real on the critical line.
For simplicity in numerical calculations, we define and use an alternative form
of the Riemann xi-function:
\begin{equation}
  \bar \xi (t) \equiv \frac{1}{\sqrt{2\pi}}\frac{t^{1/4}}{1/4+t^2} 
            e^{\frac{\pi}{4}|t|} \xi(1/2+it)
\end{equation}
because $\bar \xi (t)$, unlike $\xi(1/2+it)$, does not vanish exponentially at
large $t$.%
\footnote{Note that
    $|\Gamma(1/4+is)| \sim e^{- \pi s/2} s^{-1/4} \sqrt{2\pi}$ as $s\to0$.
}
We define
\begin{equation}
  y'(x) = \bar\xi(xy) \label{eq:zeta} 
\end{equation}
and calculate its eigenvalues and eigensolutions.

\begin{figure}
  \includegraphics[width=0.6\textwidth]{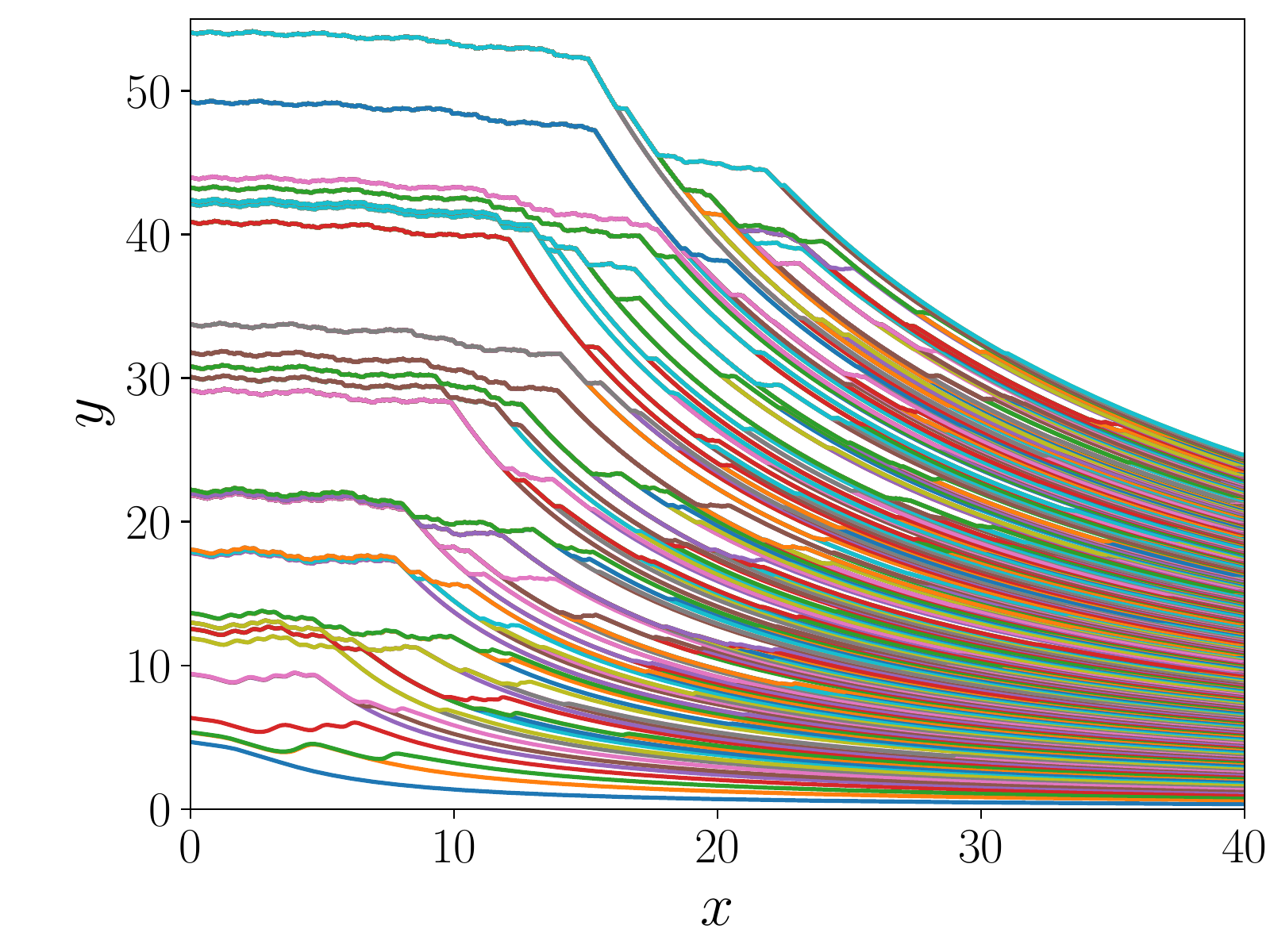}
   \caption{The first 320 eigensolutions of equation \eqref{eq:zeta}.
  }
  \label{fig:zeta_first_320}
\end{figure}

Figure \ref{fig:zeta_first_320} shows the first 320 eigensolutions of equation
\eqref{eq:zeta}. As the graph indicates, the eigenvalues obtained from this
problem inherit the quasi-random nature of the zeros of the zeta function.
One can also observe the phenomenon of
\emph{hyperfine splitting}~\cite{Bender:2019gen} between different eigenvalues.
For instance, the second and third eigenvalues form a set of eigenvalues with
hyperfine splitting; see the second (orange) and third (green) curves from below.
This problem has fascinating aspects, and we leave it to another paper.

In the above examples, we studied only first-order nonlinear eigenvalue problems.
Second-order equations, e.g., the Painlev\'e equations, provide even a richer
area of research.
Reference~\cite{Bender:2015bja} investigates the applications of nonlinear
eigenvalue problems to the first and second Painlev\'e equations and obtains
the asymptotic behavior of their eigenvalues by relating these equations to
the Schr\"odinger equation with $\mathcal{PT}$-symmetric Hamiltonian
$H = \hat p^2 + g \hat x^2 (i\hat x)^{\epsilon}$,
with $\epsilon = 1$ and 2, respectively.
References~\cite{Long:2017abc, Long:2020abc}
obtain the same results at a rigorous level for the first and second Painlev\'e
equations, respectively.
Remarkably, the large eigenvalues of the fourth Painlev\'e are also related
to the eigenvalues of the $\mathcal{PT}$-symmetric Hamiltonian with
$\epsilon = 4$~\cite{Bender:2021ngq}.
It would be interesting to extend the study to
the third, fifth, and sixth Painlev\'e equations.

Further investigations in the context of nonlinear eigenvalue problems resulted
in introducing a new class of second-order ordinary differential equations 
called generalized Painlev\'e equations~\cite{Bender:2019gen}.
Reference~\cite{Bender:2019gen} obtains these equations by loosening the
so-called \emph{Painlev\'e property} such that the movable singularities of
solutions can be either poles or fractional powers.
Given the fact that the Painlev\'e equations appear in many areas of
mathematical physics---see
references \cite{Wu:1975mw, Jimbo:1980abc, Brezin:1990rb, Douglas:1989ve,
Gross:1990abc, Moore:1990mg, Moore:1990cn, FoKas:1991za}
for a small sample---although
they were initially classified out of theoretical curiosity,
one can imagine that the generalized Painlev\'e equations find their
applications in mathematical physics too.

\acknowledgments
The author thanks Qing-hai Wang for his suggestion to investigate the reciprocal
gamma and the Riemann zeta functions.

\bibliographystyle{apsrev4-1}
\bibliography{References.bib}

\end{document}